\documentclass[a4paper]{jpconf}
\usepackage{graphicx}
\usepackage{subfigure}

\begin{document}
\title{Transverse hydrodynamics and the early-thermalization problem at RHIC}
 
\author{Radoslaw Ryblewski}
\address{The H. Niewodnicza\'nski Institute of Nuclear Physics,\\
  Polish Academy of Sciences,\\
  PL-31-342 Krak\'ow, Poland}
\ead{Radoslaw.Ryblewski@ifj.edu.pl}

\begin{abstract}
The problem of early thermalization of  matter produced in relativistic heavy-ion collisions at RHIC is discussed in the framework of a hybrid model that consists of the transverse-hydrodynamics stage followed by the standard perfect-fluid hydrodynamics stage and freeze-out. The two hydrodynamic regimes are connected with the help of Landau matching conditions. A satisfactory description of the soft hadronic observables is achieved. This indicates that the early-thermalization problem and also the HBT puzzle may be to large extent circumvented.
\end{abstract}
\section{Introduction}

The space-time evolution of matter produced in heavy-ion collisions at RHIC can be very well described by the relativistic hydrodynamics of an almost perfect fluid \cite{Kolb:2003dz, Huovinen:2003fa, Shuryak:2004cy, Teaney:2001av, Hama:2005dz, Hirano:2007xd, Nonaka:2006yn, Bozek:2009ty}. However, the use of hydrodynamics encounters the problem of early thermalization. The fast equilibration of matter assumed in the hydrodynamic picture is naturally understood in the framework of sQGP \cite{Shuryak:2004kh} or wQGP  with strong plasma instabilities \cite{Mrowczynski:2005ki}. On the other hand,  the fast thermalization is difficult to explain in the framework of the microscopic models of  early stages of collisions (e.g., in the string models). The detailed analysis of such models indicates that only transverse degrees of freedom undergo thermalization \cite{Bialas:1999zg}. In a certain time interval after the collision, the system remains highly anisotropic with the transverse pressure much larger than the longitudinal one. The calculations using this idea (the so called transverse hydrodynamics) have turned out to be consistent with the data  \cite{Bialas:2007gn,Chojnacki:2007fi,Ryblewski:2009hm,Florkowski:2009wb}. 

Despite the initial pressure asymmetries, the RHIC data suggest that the full thermalization eventually takes place. This observation has inspired us to construct a hybrid hydrodynamic model \cite{Ryblewski:2010tn} which includes the step-like transition from the transverse-hydrodynamics stage to  the perfect-fluid hydrodynamics stage. 

We assume that the system formed after the collision undergoes only transverse thermalization and in the proper time interval $\tau_{\rm i} \leq \tau \leq \tau_{\rm tr} $ can be described by the tranverse hydrodynamics. The time $\tau_{\rm i}$ is the initialization time for transverse hydrodynamics, whereas $\tau_{\rm tr}$ is the transition time to the perfect-fluid hydrodynamics. During this transition the energy  and momentum are conserved. Additionally, we require that the entropy increases. After the transition, for $\tau > \tau_{\rm tr} $, the system is treated as the perfect fluid. The single freeze-out model is applied on the hypersurface of constant temperature $T_{3f}$ to calculate physical observables. The hydrodynamic code is coupled to the statistical hadronization Monte-Carlo model {\tt THERMINATOR} \cite{Kisiel:2005hn}, which is used to calculate the soft hadronic observables: the transverse-momentum spectra of pions, kaons, and protons, the elliptic flow of pions+kaons and protons, and the HBT radii of pions. The radii are calculated with the method described in \cite{Kisiel:2006is}.

The idea of the initial transverse-hydrodynamics stage is used to circumvent the problem of very early thermalization. There exists, however, another problem connected with the hydrodynamic picture. The so called HBT puzzle refers to a difficulty of simultaneous correct description of the hadronic transverse-momentum spectra, the elliptic flow coefficient, and the Hanbury-Brown--Twiss interferometry data. Interestingly, the inclusion of the transverse-hydrodynamics stage improves the HBT results \cite{Ryblewski:2010tn}.

\begin{figure}[t]
\begin{center}
\subfigure{\includegraphics[angle=0,width=0.65\textwidth]{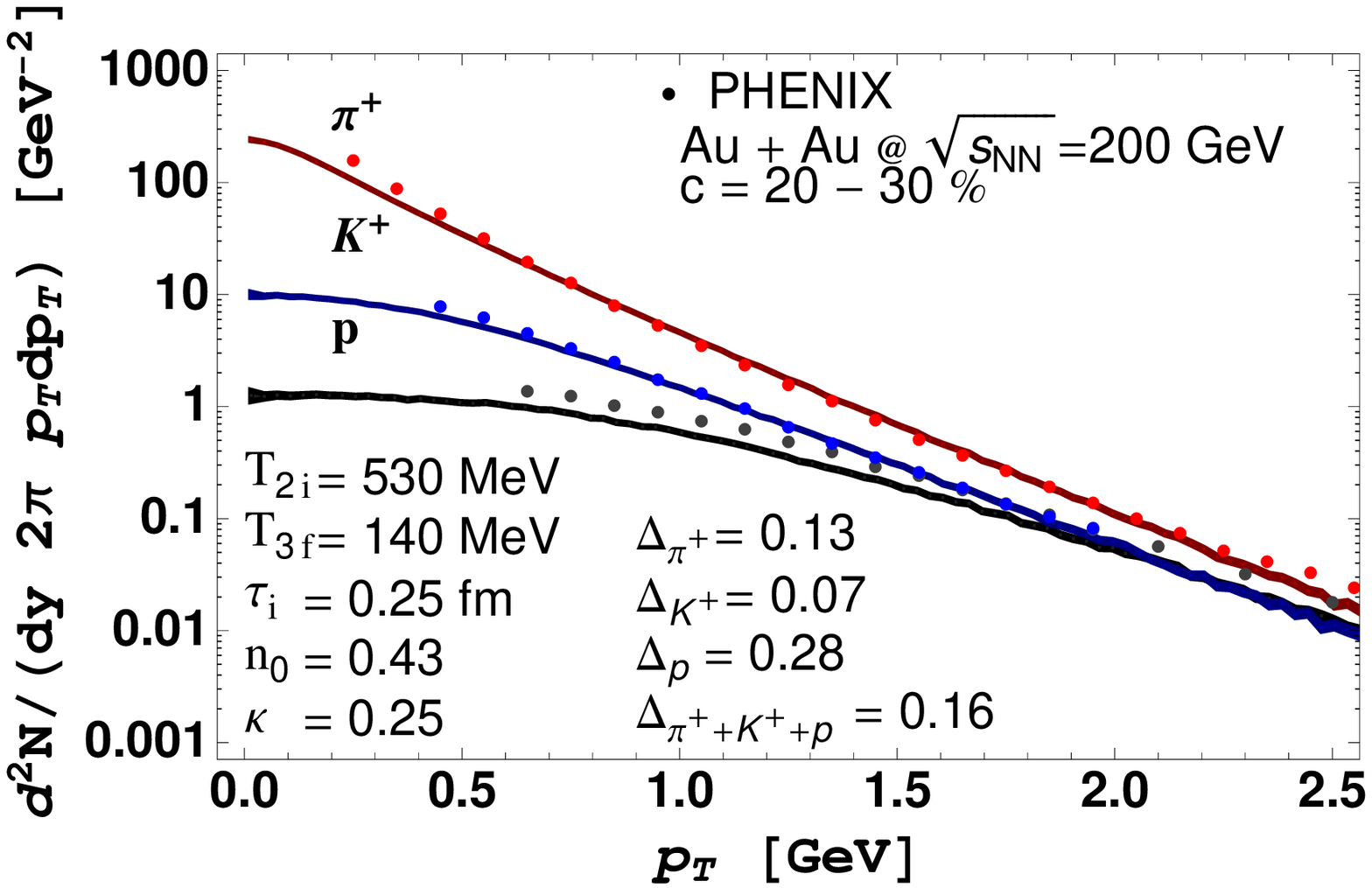}} \\
\vspace{1cm}
\subfigure{\includegraphics[angle=0,width=0.65\textwidth]{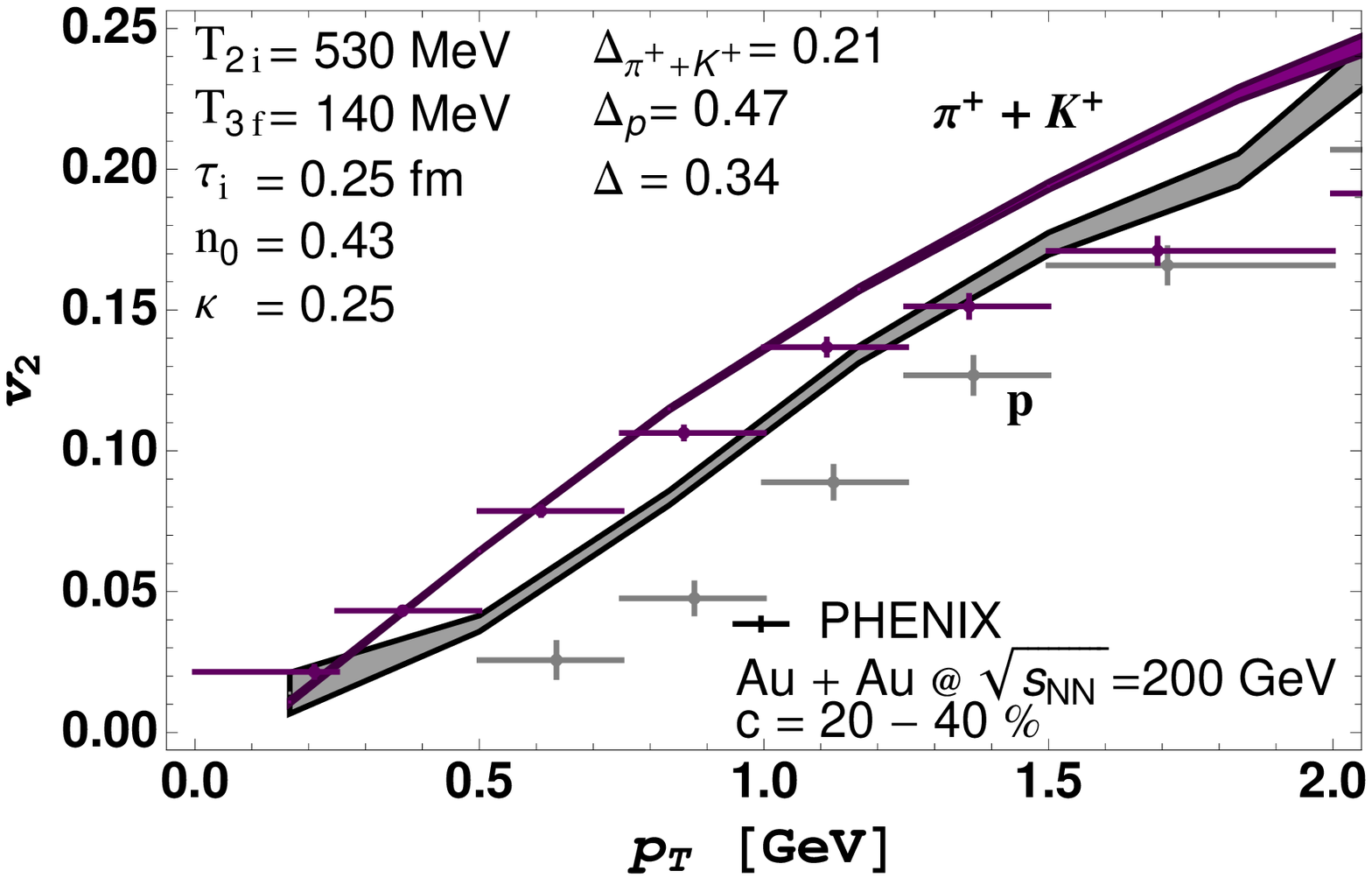}} 
\end{center}
\caption{(Color online) Upper part: The comparison of the model (lines) and experimental  (dots) transverse-momentum spectra of pions, kaons, and protons. The data are taken from Ref. \cite{Adler:2003cb}. Lower part: The comparison of the model and experimental  elliptic flow coefficient $v_2$ for pions+kaons and protons. The bands indicate the error of the Monte-Carlo calculations. The data are taken from Ref. \cite{Adler:2003kt}.}
\label{fig:sp}
\end{figure}

\section{Results}

In the upper part of Fig. \ref{fig:sp} we show our model results for the transverse-momentum spectra of pions, kaons and protons compared to the PHENIX data \cite{Adler:2003cb} for Au+Au collisions at the energy of $\sqrt{s_{NN}}$ = 200 GeV and for the centrality class 20-30\%. One observes overall good agreement between the model predictions and the data, at the level of 16\%. The model results have been obtained with the initial central temperature $T_{2i}$  = 530 MeV and the freeze-out temperature $T_{3f}$ = 140 MeV. The fraction of binary collisions in the initial profile of the density of sources of particle production is 0.25 and the density of transverse clusters in rapidity $n_0$ = 0.43 (our fitting procedure is described in more detail in \cite{Ryblewski:2010tn}). 

In the lower part of the Fig. \ref{fig:sp} we show our results for the elliptic flow coefficient $v_2$ compared with the PHENIX \cite{Adler:2003kt} experimental data. The pion+kaon data are reproduced at the level of 21\% while the model results for the protons overestimate the data by about 50 \%. The reason for too large proton $v_2$ is the lack of hadron rescattering in the final state and/or the lack of viscous effects.

\begin{figure}[t]
\begin{center}
\subfigure{\includegraphics[angle=0,width=0.65\textwidth]{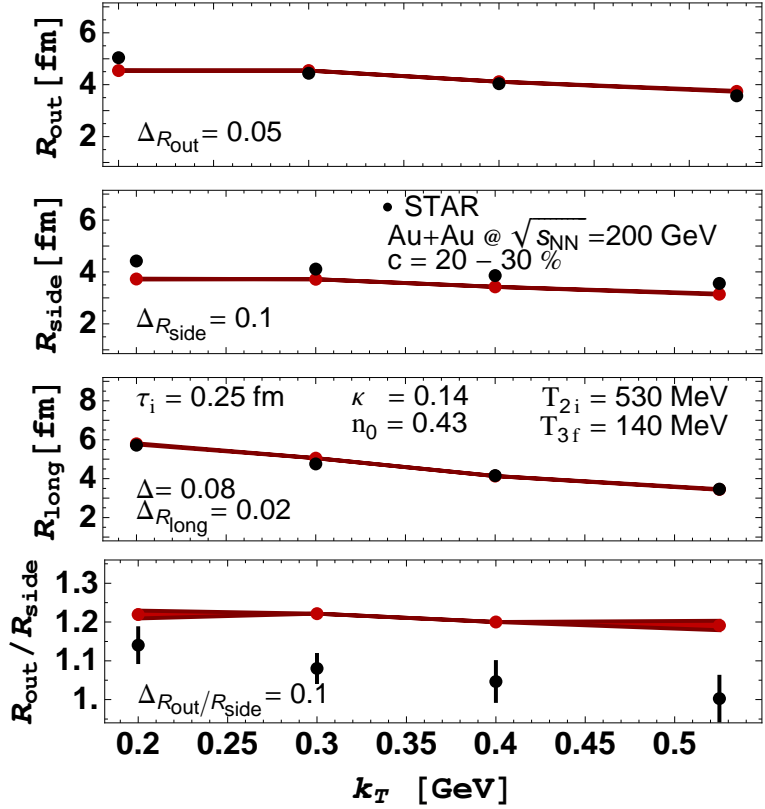}}
\end{center}
\caption{(Color online) The comparison of the model (red lines) and experimental (black dots) results for the pionic HBT radii. The data are taken from Ref. \cite{Adams:2004yc}. The model parameters the same as in Fig. \ref{fig:sp}. }
\label{fig:hb}
\end{figure}

     In Fig. \ref{fig:hb} we present the model HBT radii compared to the STAR \cite{Adams:2004yc} data. We observe very good agreement between the experimental data and the results taken from the model analysis.  All radii are reproduced with the accuracy better than 10\%.

\section{Conclusions}

Our analysis shows that it is possible to obtain a satisfactory description of the soft hadronic RHIC data in a model that does not assume a very fast thermalization of the system. Initially, only transverse degrees of freedom are thermalized, as suggested by the particle production from the fluctuating strings. Our model reproduces well the HBT radii as well as other soft hadronic observables. This indicates that the two main RHIC problems may be to large extent circumvented. The proton elliptic flow that is too large may be corrected by the inclusion of the shear and bulk viscosity in the standard hydrodynamic stage or by the inclusion of the hadronic rescattering after freeze-out. 

Recently we have also analyzed a physical scenario where the transverse hydrodynamics was followed by a sudden isotropization and freeze-out \cite{Ryblewski:2009hm,Florkowski:2009wb}. The present model differs from \cite{Ryblewski:2009hm,Florkowski:2009wb} by the inclusion of the extended phase described by the perfect-fluid hydrodynamics. The similarities between our present results and the results of Refs. \cite{Ryblewski:2009hm,Florkowski:2009wb} indicate that the details of modeling the equilibration transition might be not very much important. However, the correct description of the HBT radii is achieved only if the full equilibration of the system is reached. 

The present approach may be generalized in several ways. In particular, the perfect-fluid stage may be replaced by the viscous hydrodynamics. The viscous effects may be also included in the transverse hydrodynamics. Naturally, the next step should be a construction of a continuous change from transverse hydrodynamics to perfect-fluid hydrodynamics (first steps in this direction have been done very recently in  \cite{Florkowski:2010cf}). 

Finally we note that the large initial anisotropy of pressure may be explained only by introducing a very large shear viscosity, hence our model describes the initial processes which are out of range of application of the standard viscous hydrodynamics.

\end{document}